\documentclass[aps,twocolumn]{revtex4}
\usepackage{hyperref}
\usepackage[sort&compress]{natbib}
\usepackage{amsfonts}
\usepackage{amsmath}
\usepackage{amssymb,epsf}
\usepackage{latexsym}
\usepackage{graphicx,epsfig}
\usepackage{epstopdf}
\usepackage{caption}
\graphicspath{{./figs/}}

\begin{document}
\title{Ghost Dark Energy in the Modified Kaniadakis Cosmology}

\author{A. Dezhakam}
\author{A. Sheykhi}
\email{asheykhi@shirazu.ac.ir}
\author{A. Dehyadegari}
\email{amin.dehyadegari@hafez.shirazu.ac.ir}

\address{Department of Physics, College of
Science, Shiraz University, Shiraz 71454, Iran and\\
Biruni Observatory, College of Science, Shiraz University, Shiraz
71454, Iran}

\begin{abstract}

We investigate ghost dark energy (GDE) in a cosmological framework
derived from Kaniadakis entropy. By applying the first law of
thermodynamics to the FRW apparent horizon, we obtain modified
Friedmann equations that include a correction term characterized
by the Kaniadakis parameter $\lambda$. We then study the evolution
of a flat universe containing pressureless matter and interacting
GDE within this modified gravity setup. Our numerical analysis
reveals that the Kaniadakis correction mildly affects the dark
energy equation of state and shifts the transition to cosmic
acceleration. Stability analysis via the squared sound speed shows
the model remains generally unstable, though the instability is
moderated for larger $\lambda$. Statefinder diagnostics indicate
that the model approaches the $\Lambda$CDM fixed point in the
future, with deviations decreasing as $\lambda$ increases.

\end{abstract}

\maketitle
\newpage

\section{Introduction}\label{intro}
The discovery of the accelerated expansion of the universe from
Type Ia supernova observations~\cite{riess1998,perlmutter1999},
later confirmed by cosmic microwave background (CMB) and
large-scale structure (LSS) data~\cite{spergel2003,tegmark2004},
has established the existence of a mysterious component known as
dark energy (DE) that drives this cosmic acceleration. Despite
significant observational evidence, the fundamental nature of dark
energy remains one of the greatest unsolved problems in modern
cosmology. The simplest and most successful phenomenological model
is the cosmological constant $\Lambda$, which corresponds to a
constant energy density and an equation of state (EoS) parameter,
$w = -1$. While the $\Lambda$CDM model provides an excellent fit
to current observations, it suffers from severe theoretical
issues, including the fine-tuning and coincidence
problems~\cite{weinberg1989,peebles2003}. These difficulties have
motivated extensive efforts to explore alternative dynamical dark
energy models, such as
quintessence~\cite{ratra1988,wetterich1988},
phantom~\cite{caldwell2002,nojiri2011}, and holographic dark
energy~\cite{mli2004,nojiri2006,wang,Em1}, among others.

Among various dynamical dark energy models, GDE has attracted
considerable attention due to its solid theoretical foundation in
quantum chromodynamics (QCD). The GDE model arises from the
Veneziano ghost field in QCD, which is known to contribute a small
vacuum energy density of order $\Lambda_{\text{QCD}}^3 H$, where
$\Lambda_{\text{QCD}} \sim 100$ MeV is the QCD mass scale and $H$
is the Hubble parameter~\cite{urban2010a,urban2010b}. Remarkably,
this energy density is of the same order as the observed dark
energy density, naturally alleviating the fine-tuning problem
without introducing new degrees of freedom. In an expanding
universe, the GDE energy density takes the simple form $\rho_D =
\beta H$, where $\beta$ is a constant parameter related to the QCD
scale~\cite{cai2011notes}. This model has been extensively studied
in various cosmological contexts and has been shown to be
consistent with observational
data~\cite{cai2012more,malekjani2013}.

On the other hand, there is growing evidence suggesting a deep
connection between gravity and thermodynamics. This line of
thought was pioneered by the seminal works of
Jacobson~\cite{jacobson1995}, who derived the Einstein field
equations from the Clausius relation applied to local Rindler
horizons, and by
Padmanabhan~\cite{padmanabhan2005,padmanabhan2010}, who emphasized
the thermodynamic origin of gravitational dynamics. Building upon
these ideas, Cai and Kim~\cite{cai2005,wang2,Cai2} showed that the
Friedmann equations can be derived from the first law of
thermodynamics applied to the apparent horizon of an FRW universe,
assuming the standard Bekenstein-Hawking entropy $S = A/4G$. This
profound correspondence has led to the development of
entropy-modified cosmologies, where deviations from the standard
area law lead to modified Friedmann equations and potentially new
cosmological dynamics.

In this context, the Kaniadakis
entropy~\cite{kaniadakis2002,kaniadakis2005} represents a
relativistic generalization of the Boltzmann-Gibbs entropy,
arising from a consistent relativistic statistical framework. The
Kaniadakis entropy is characterized by a deformation parameter $K$
and reduces to the standard entropy in the limit $K \to 0$. This
entropy has been successfully applied to various physical systems,
including black holes and
cosmology~\cite{abreu2016,abreu2017,abreu2018,abreu2019}. Modified
Friedmann equations have been derived, by applying the Kaniadakis
entropy to the apparent horizon \cite{sheykhi2024}. Further
cosmological implications of Kaniadakis entropy have been explored
in recent works~\cite{lymperis2021,luciano2022,Shey2,dehyadegari2026modified}.

Motivated by these developments, in this work we investigate the
cosmological evolution of GDE within the framework of Kaniadakis
entropy-modified gravity. Specifically, we consider a spatially
flat FRW universe filled with pressureless matter and interacting
GDE, and derive the modified Friedmann equations from the
Kaniadakis entropy. We then study the dynamics of the model by
numerically solving the evolution equations for the GDE density
parameter and the Hubble parameter. We analyze the effects of the
Kaniadakis correction parameter and the matter-DE interaction on
the dark energy equation of state, the deceleration parameter, and
the cosmic transition from deceleration to acceleration.
Furthermore, we examine the stability of the model against
perturbations using the squared sound speed and employ the
statefinder diagnostic to distinguish our model from the standard
$\Lambda$CDM paradigm.

The structure of the paper is as follows. In Section II, we
briefly review the Kaniadakis cosmology and derive the modified
Friedmann equations from the thermodynamic-gravity conjecture. In
Section III, we study the GDE model in Kaniadakis cosmology, and
derive the evolution equations for the density parameter and EoS
parameter in the presence of interaction. Section IV presents the
numerical analysis and discusses the cosmological consequences,
including stability and statefinder diagnostics. Finally, Section
V contains our closing remarks and conclusions.
\section{Kaniadakis cosmology}\label{Kaniadakis}
In this section, we construct a cosmological framework based on
Kaniadakis entropy, motivated by the deep connection between
thermodynamics and gravity. This is achieved through a
modification of the standard Bekenstein-Hawking entropy, leading
to the corresponding modified Friedmann equations. For further
details on Kaniadakis entropy, see Refs.
\cite{abreu2016,abreu2017,abreu2018,abreu2019}.

We assume a spatially homogeneous and isotropic FRW background described by the line element
\begin{equation}\label{ds}
ds^2=h_{\mu\nu}dx^{\mu}dx^{\nu}+\tilde{r}^2(d\theta^2+sin^2\theta
d\phi^2),
\end{equation}
where $x^0=t,x^1=r, \tilde{r}=a(t)r$ and $a(t)$ denotes the scale
factor. Here $h_{\mu\nu} = \mathrm{diag}(-1, a^2(t)/[1-kr^2])$ is
the two-dimensional spacetime metric, where $k =  0,-1, +1$
characterizes  flat, open and closed spatial geometries,
respectively. From a thermodynamic viewpoint, the apparent horizon
provides a natural boundary where the laws of thermodynamics are
satisfied. The radius of the apparent horizon is given by
\cite{Hay1,Hay2,Bak}
\begin{equation}\label{app}
\tilde{r}_{h}=\frac{1}{\sqrt{H^2+{k}/{a^2}}},
\end{equation}
where $ H=\dot{a}/a$ is the Hubble parameter and an overdot
indicates the derivative with respect to $t$. The temperature
associated with the horizon takes the form
\cite{Hay1,Hay2,Bak,Cai2}
\begin{equation}\label{temp}
T_h=-\frac{1}{2\pi\tilde{r}_h}\left(1-\frac{\dot{\tilde{r}}_h}{2H\tilde{r}_h}\right).
\end{equation}
The matter content of the FRW universe is assumed to be a perfect
fluid, whose energy-momentum tensor can be written as
\begin{equation}\label{tmn}
T_{\mu\nu}=(\rho+p)u_{\mu}u_{\nu}+pg_{\mu\nu},
\end{equation}
where $u_{\mu}=(1,0,0,0)$ is the four-velocity of the comoving
observer, satisfying the condition $u_{\mu}u^{\mu}=1$, and $\rho$
and $p$ are the energy density and pressure of the fluid,
respectively. By imposing the conservation of the energy-momentum
tensor, $\nabla_\mu T^{\mu\nu}=0$, we obtain the following
relation between the energy density and pressure:
\begin{equation}\label{coneq}
\dot{\rho}+3H(\rho+p)=0.
\end{equation}
With the energy-momentum tensor at hand, we can define the work
density in an expanding universe as \cite{Hay1}
\begin{equation}\label{work}
W=-\frac{1}{2}T^{\mu\nu}h_{\mu\nu},
\end{equation}
which in terms of the energy density and pressure takes the form
\begin{equation}\label{workEP}
W=\frac{1}{2}(\rho-p).
\end{equation}
Now, by applying the thermodynamics-gravity conjecture at the
apparent horizon, we derive the Friedmann equations from the laws
of thermodynamics. Accordingly, we write the first law of
thermodynamics at the apparent horizon as
\begin{equation}\label{flaw}
dE=T_h\,dS_h+WdV_h,
\end{equation}
where $W$ is the work density, $T_h$ and $S_h$ are the temperature
and entropy of the horizon, respectively. Moreover, $E$ and $V_h$
denote the total energy of the universe and the volume of a sphere
with radius $\tilde{r}_h$, namely,
\begin{equation}\label{EnergyVolume}
E=\rho V_h, \qquad V_h=\frac{4\pi}{3}\tilde{r}_h^3.
\end{equation}
By differentiating the above total energy and making use of the
conservation equation \eqref{coneq}, we arrive at
\begin{equation}\label{der}
dE=4\pi \tilde{r}^2_h \rho d\tilde{r}_h-4\pi H \tilde{r}^3_h(\rho+p)dt.
\end{equation}
In the following, we adopt the modified Kaniadakis entropy for the
apparent horizon, given by
\begin{equation} \label{SK2}
S_h=S_{K}=\frac{1}{K}\sinh{(K S)},
\end{equation}
where $K$ is the Kaniadakis parameter, and $S$ is the standard
Bekenstein-Hawking entropy, defined in terms of the apparent
horizon area and the gravitational constant as
\begin{equation}
S=\frac{A}{4G} =\frac{\pi\tilde{r}_{h}^2}{G}.
\end{equation}
If we expand expression (\ref{SK2}) for $K\ll1$, we arrive at
\cite{sheykhi2024}
\begin{equation}
S_{h}=S+\frac{K^2}{6}S^3.
\end{equation}
The differential form of the Kaniadakis entropy is given by
\begin{equation}
dS_{h}=\left(1+\frac{K^2}{2}S^2\right)\,dS,
\end{equation}
in which
\begin{equation}
 dS=\frac{2\pi \tilde{r}_{h}{}}{G}\dot{\tilde{r}}_h\, dt.  \notag
\end{equation}
Upon substituting the entropy expression above along with the
horizon temperature \eqref{temp}, work density \eqref{workEP}, and
volume \eqref{EnergyVolume} into the first law of thermodynamics
\eqref{flaw}, together with the continuity equation \eqref{coneq},
we obtain the following differential relation (see
\cite{sheykhi2024} for details),
\begin{equation}
\left(1+\frac{K^{2}\pi ^{2}}{2G^{2}}\tilde{r}_{h}^4\right) \frac{d\tilde{r}_{h}}{\tilde{r}_{h}^{3}}=-%
\frac{4\pi G}{3}d\rho .
\end{equation}
Integrating the above relation and using the connection between
the Hubble parameter and the apparent horizon radius \eqref{app},
the modified Friedmann equation for a spatially flat universe
takes the following form \cite{sheykhi2024}
\begin{equation} \label{FFEq}
 H^2-\alpha H^{-2}=\frac{8\pi G}{3}(\rho+{\rho_{\Lambda}}),
\end{equation}
where
\begin{equation}
 \alpha=\frac{K^2 \pi ^2}{2G^2},\qquad \rho_{\Lambda}=\frac{\Lambda}{8 \pi G}, \notag
\end{equation}
with $\Lambda$ being an integration constant, naturally identified
as the cosmological constant.

In the next section, we investigate the cosmological evolution of
the universe within the framework of the modified Friedmann
equation derived above \eqref{FFEq}, where we also take into
account an additional dark energy component.
\section{Ghost dark energy in Kaniadakis cosmology} \label{GDEKaniadakis}
Now, we study the cosmological evolution of a spatially flat
universe filled with pressureless matter and GDE within the
framework of Kaniadakis-modified cosmology, allowing for a
possible interaction between the two components. To this end, we
write the first modified Friedmann equation as follows
\begin{equation}\label{1MFEq}
H^2-\alpha H^{-2}=\frac{8\pi G}{3}(\rho_{m}+\rho_{D}),
\end{equation}
where ${\rho_{m} }$ and ${\rho_{D}}$ represent the energy
densities of pressureless matter and GDE, respectively. In this
regard, the continuity equations governing both components are
given by
\begin{eqnarray}\label{conseq1}
&&\dot{\rho}_{D}+3H\rho_{D}(1+w_{D})=-Q,\\
&&\dot{\rho}_{m}+3H\rho_{m}=Q, \label{conseq2}
\end{eqnarray}
where $w_{D}=p_{D}/\rho_{D}$ is the equation-of-state parameter of
GDE, while $Q$ denotes the interaction term between the matter and
GDE components. The interaction term ($Q$) is generally introduced
phenomenologically, and several functional forms have been
proposed in the literature. The simplest and most commonly adopted
class of interaction terms is proportional to $H \rho_{cr}$, where
${\rho_{cr}}={\rho_{m}+{\rho_{D}}}$. Accordingly, we assume the
interaction term to be given by $Q=3b^2 H \rho_{cr}$, where $b$ is
a dimensionless coupling constant that characterizes the strength
of the interaction.

One of the main objectives of this work is to study the GDE model
in entropy-corrected cosmological frameworks. In an expanding
universe, the energy density of GDE is described by
\cite{cai2011notes}
\begin{equation} \label{EDGDE}
\rho_{D}=\beta H,
\end{equation}
where $\beta$ is a constant parameter associated with the vacuum
energy contribution of the QCD field in a dynamical universe.
Differentiating the modified Friedmann equation \eqref{1MFEq} with
respect to cosmic time yields
\begin{align}
\frac{\dot{H}}{H}&=\frac{4\pi G\dot{\rho}_{cr}}{3(H^{2}-\alpha H^{-2})}\left(
\frac{H^{2}+\alpha H^{-2}}{H^{2}-\alpha H^{-2}}\right) ^{-1}, \notag \\
&=\frac{\dot{\rho }_{cr}}{2\rho _{cr}}\left( 1+\frac{2\alpha
H^{-2}}{H^{2}-\alpha H^{-2}}\right) ^{-1},
\end{align}
where the second line follows from substituting Eq. \eqref{1MFEq}.
Making use of the continuity equations \eqref{conseq1} and
\eqref{conseq2} together with the GDE energy density
\eqref{EDGDE}, the above relation can be written in the following
form
\begin{equation}\label{OHdot}
\frac{\dot{H}}{H}=-\frac{3}{2}H(1+w_{D} \, \Omega _{D})\left( 1+\frac{\lambda \, \Omega _{D}}{H^{3}}\right) ^{-1},
\end{equation}
where $\lambda =3\alpha /(4\pi \beta G  )$ and $\Omega _{D}$
represents the fractional energy density of GDE,
$\Omega_D=\rho_D/\rho_{cr}$. It is worth noting that the parameter
$\lambda$ characterizes the contribution of the Kaniadakis entropy
correction, such that when $\lambda=0$, the standard GDE results
are recovered \cite{cai2011notes}. In the limit of small
deviations from the standard entropy, the above relation
\eqref{OHdot} reduces to
\begin{equation}\label{Hdot}
\frac{\dot{H}}{H}=-\frac{3}{2}H(1+w_{D} \, \Omega _{D})\left( 1-\frac{\lambda \, \Omega _{D}}{H^{3}}\right).
\end{equation}
By combining the continuity equation for GDE \eqref{conseq1} with
the GDE energy density expression \eqref{EDGDE}, the interaction
term $Q$, and the relation \eqref{Hdot}, one finds
\begin{equation}\label{omegaD}
w_{D}=-\frac{2+2 \, b^{2}/\Omega _{D}-\left( 1-\lambda \, \Omega _{D}/H^{3}\right)}{2-\Omega
_{D}\left( 1-\lambda \, \Omega _{D}/H^{3}\right)}.
\end{equation}
Upon differentiating $\Omega _{D}$ with respect to cosmic time and applying the first modified Friedmann equation \eqref{1MFEq}, we arrive at
\begin{equation}
\frac{\dot{\Omega}_{D}}{\Omega _{D}}=-\frac{\dot{H}}{H}\left( 1+2\frac{\lambda \, \Omega
_{D}}{H^{3}}\right) ,
\end{equation}
Introducing ``$\ln a$'' as the new independent variable and
utilizing the relations \eqref{Hdot} and \eqref{omegaD}, the above
expression takes the following form
\begin{equation}\label{OmegaD}
\frac{\Omega _{D}^{\prime }}{\Omega _{D}}=-\frac{3\left( 1-\Omega
_{D}-b^{2}\right) (1-\lambda \, \Omega _{D}/H^{3})}{2-\Omega _{D}(1-\lambda \,
\Omega _{D}/H^{3})}\left( 1+2\frac{\lambda \, \Omega _{D}}{H^{3}}\right) ,
\end{equation}
where the prime indicates differentiation with respect to $\ln a$.
As expected, setting $\lambda=0$ reproduces the standard GDE
evolution equation \cite{cai2011notes}. To solve the evolution
equation of $\Omega _{D}$, the Hubble parameter $H$ is required,
which can be obtained from the first modified Friedmann equation
\eqref{1MFEq}. To this end, using the GDE energy density given in
\eqref{EDGDE}, the first modified Friedmann equation takes the
following form
\begin{equation}\label{MFFN}
\frac{3H^{2}}{4\pi \beta G}-\lambda \, H^{-2}-\frac{2H}{\Omega _{D}}=0.
\end{equation}
By simultaneously solving Eqs. \eqref{OmegaD} and \eqref{MFFN}, we
obtain the evolution of the Hubble parameter and $\Omega _{D}$,
which allows us to examine the cosmological consequences of the
Kaniadakis entropy model. One of the important quantities
characterizing cosmic evolution is the deceleration parameter,
defined as
\begin{equation}
q=-1-\frac{\dot{H}}{H^2},
\end{equation}
which, upon substituting relations \eqref{Hdot} and \eqref{omegaD}, can be expressed as
\begin{equation}
q=-1+3\frac{\left( 1-\Omega _{D}-b^{2}\right) \left( 1-\lambda \, \Omega _{D}/H^{3}\right)}{%
2-\Omega _{D}\left( 1-\lambda \, \Omega _{D}/H^{3}\right)}.
\end{equation}
It is worth mentioning that when $\lambda=0$, the standard GDE
deceleration parameter is recovered \cite{cai2011notes}. In the
following section, we numerically solve the governing equations
and discuss the resulting cosmological evolution of the model.
\section{Numerical Analysis of Kaniadakis GDE (KGDE) Model}\label{Numerical}
In this section, we numerically solve Eqs. \eqref{OmegaD} and
\eqref{MFFN} to investigate the effects of the Kaniadakis
parameter ($\lambda$) and the interaction term ($b^{2}$) on the
cosmological evolution of the universe. To this aim, we define a
dimensionless version of $\lambda$ as $\lambda_r=\lambda/H_0^3$,
and perform the numerical analysis for small values of
$\lambda_r$. In addition, we set $\Omega _{D0}=0.69$ and
$\beta=0.25H_0/(\pi G )$ in our numerical analysis. It should be
noted that the values of $\lambda_r$ used in the numerical
analysis are consistent with the small deviation approximation
employed in deriving Eq. \eqref{Hdot}.

\begin{figure}
\centering
\includegraphics[scale=0.9]{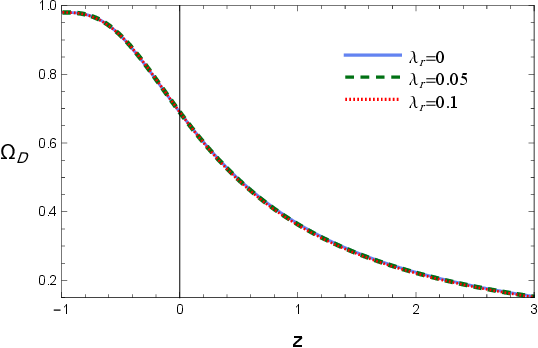}
\includegraphics[scale=0.9]{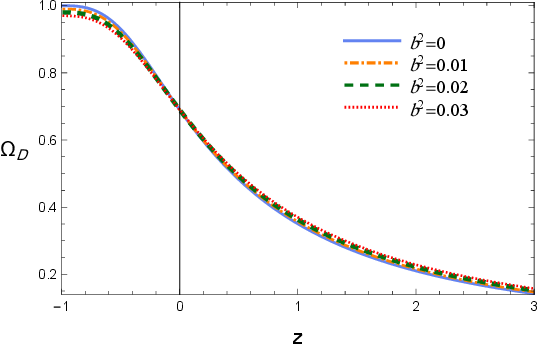}
\vspace{2mm}
\caption{Evolution of $\Omega_{D}$ parameter for GDE in Kaniadakis cosmology. In upper
panel we set $b^2 = 0.02$ while in lower panel we take  $\lambda_r=0.05$.} \label{fig:OmegaD}
\end{figure}

\begin{figure}
\centering
\includegraphics[scale=0.9]{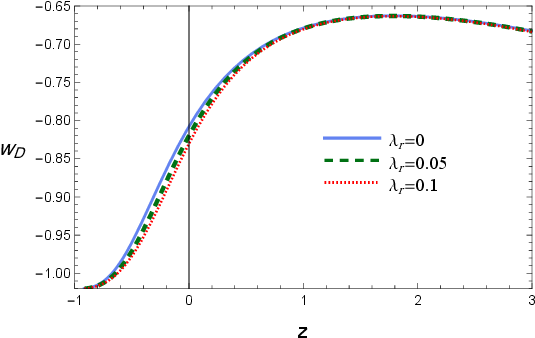}
\includegraphics[scale=0.9]{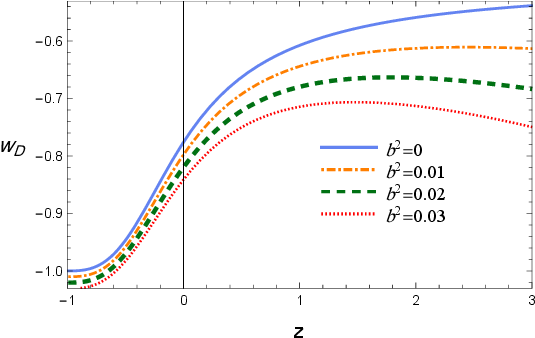}
\vspace{2mm}
\caption{Evolution of $w_D$ parameter for GDE in Kaniadakis cosmology. In upper
panel we set $b^2 = 0.02$ while in lower panel we take $\lambda_r=0.05$.}
\label{fig:wD}
\end{figure}

\begin{figure}
\centering
\includegraphics[scale=0.9]{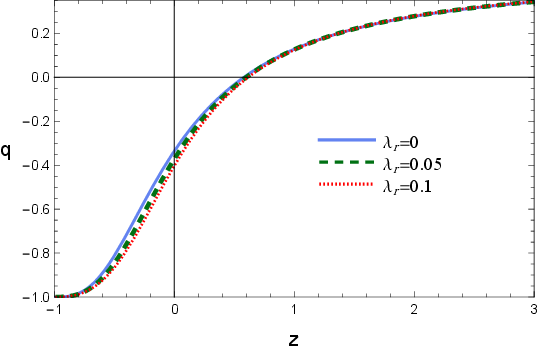}
\includegraphics[scale=0.9]{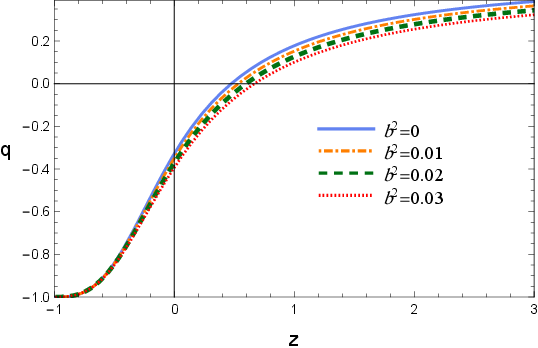}
\vspace{2mm}
\caption{ Evolution of $q$ parameter for GDE in Kaniadakis cosmology. In upper
panel we set $b^2 = 0.02$ while in lower panel we take $\lambda_r = 0.05$.} \label{fig:q}
\end{figure}

The evolution of $\Omega _{D}$, $w _{D}$, and $q$ as a function of
redshift $z$ for different values of $\lambda_r$ and $b^{2}$ are
depicted in Figs. \ref{fig:OmegaD}-\ref{fig:q}. As can be seen
from Fig. \ref{fig:OmegaD}, the GDE density parameter $\Omega
_{D}$ is negligibly small at early times, while the model predicts
a dark-energy-dominated universe in the future. The upper panel of
Fig. 1 shows that, in the presence of interaction $b^{2}=0.02$,
the Kaniadakis parameter $\lambda_r$ has no significant effect on
the evolution of $\Omega _{D}$. The lower panel illustrates the
impact of the interaction parameter $b^{2}$ on the evolution of
$\Omega _{D}$ in the presence of the Kaniadakis parameter
$\lambda_r=0.05$. From this panel, one can observe that increasing
the interaction between matter and the dark energy component
slightly enhances $\Omega _{D}$ at early times, while it leads to
a slight decrease in the future. Fig. 2 shows the behavior of the
dark energy equation-of-state parameter $w _{D}$ as a function of
redshift $z$.
According to the upper panel, the Kaniadakis parameter $\lambda_r$
has no significant effect on $w _{D}$ at early and future times,
while it slightly affects its behavior around the present time. In
particular, around the present epoch, $w _{D}$ shows a decreasing
trend with increasing $\lambda_r$. The lower panel shows that the
interaction parameter $b^2$ significantly influences $w _{D}$ at
high redshifts. Moreover, within the Kaniadakis cosmological
framework, the inclusion of the interaction term allows the GDE
model to cross the phantom divide ($w_D = -1$) at late times, a
behavior that is absent in the non-interacting case. At late
times, $w _{D}$ approaches a common limiting value that is
independent of $\lambda_r$, but depends on the interaction
parameter $b^2$.

The evolution of the deceleration parameter $q$ as a function of
redshift $z$ is depicted in Fig. \ref{fig:q}. The upper panel
demonstrates that increasing $\lambda_r$ slightly shifts the
transition from decelerated to accelerated expansion toward higher
redshifts. Similarly, the lower panel shows that, for a fixed
$\lambda_r=0.05$, a stronger interaction between the dark sector
and pressureless matter shifts the transition toward higher
redshifts. In both panels, the deceleration parameter
asymptotically approaches $q=-1$ at late times, indicating a
de-Sitter expansion phase. In what follows, we examine the sound
stability and statefinder of the KGDE model.
\subsection{ Stability}
A fundamental requirement for any dark energy model is the
stability of the cosmic fluid against small perturbations. The
squared sound speed $v^2_{s}$ provides a convenient way to examine
the behavior of perturbations and assess the stability of the
model. In particular, a positive value of $v^2_{s}$ corresponds to
stable perturbations, whereas a negative value indicates the
presence of instabilities. To this end, the adiabatic squared
sound speed is defined as
\begin{equation}
    v^2_{s}=\frac{dp_D}{d\rho_{cr}},
\end{equation}
where $p_D$ is the pressure of the GDE component and $\rho_{cr}$
represents the total energy density of the cosmic fluid. Here, we
have used the fact that pressureless matter does not contribute to
the total pressure of the cosmic fluid. Based on the definition of
the equation-of-state parameter, the squared sound speed can be
expressed as
\begin{equation}
v^2_{s}=\frac{\dot{p}_D}{\dot{\rho}_{cr}}=\frac{\dot{\rho}_D}{\dot{\rho}_{cr}}(w
_{D}+\dot{w} _{D}\frac{\rho_D}{\dot{\rho}_{D}}).
\end{equation}
\begin{figure}
\centering
\includegraphics[scale=0.9]{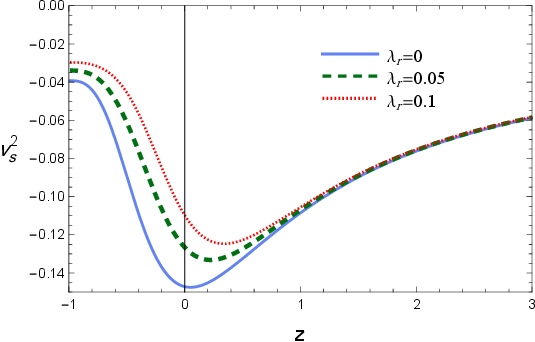}
\includegraphics[scale=0.9]{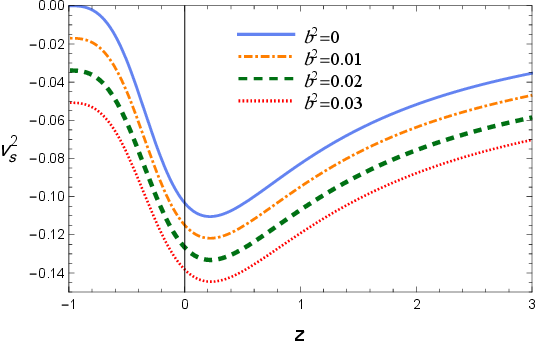}
\vspace{2mm}
\caption{The behavior of the squared sound speed $v^2_{s}$ vs $z$ in Kaniadakis cosmology. In upper
panel we set $b^2 = 0.02$ while in lower panel we take $\lambda_r = 0.05$.} \label{fig:vs2}
\end{figure}
By using the first modified Friedmann equation \eqref{1MFEq}
together with the continuity equations \eqref{conseq1} and
\eqref{conseq2}, and employing Eqs. \eqref{OHdot}, \eqref{omegaD}
and \eqref{OmegaD}, the squared sound speed can be obtained in the
following form:
\begin{align}
v_{s}^{2}& =\frac{\left( 1-\lambda \,\Omega _{D}/H^{3}\right) }{\left(
2-\Omega _{D}+\lambda \,\Omega _{D}^{2}/H^{3}\right) ^{2}}\left[
3b^{2}\Omega _{D}\left( 1-\lambda \,\Omega _{D}/H^{3}\right) \right.   \notag
\\
& \left. -4b^{2}\left( 1+\lambda \,\Omega _{D}/H^{3}\right) -\Omega
_{D}(1-\Omega _{D})\left( 1-3\lambda \,\Omega _{D}/H^{3}\right) \right] .
\end{align}
It is worth noting that, in the absence of the Kaniadakis
parameter ($\lambda=0$), the above expression reduces to the
corresponding result in standard GDE cosmology
\cite{cai2011notes}. The dependence of $v^2_{s}$ on redshift $z$
for different values of $\lambda_r$ and $b^2$ is shown in Fig.
\ref{fig:vs2}. From Fig. \ref{fig:vs2}, it is clear that the KGDE
model remains unstable under perturbations for all considered
values of $\lambda_r$ and $b^2$. Nevertheless, in the
non-interacting case ($b^2=0$) and for $\lambda_r=0.05$, the lower
panel reveals that the model approaches the stability boundary at
late times. According to the upper panel, increasing $\lambda_r$
makes $v^2_{s}$ less negative around the present and future
epochs, whereas the lower panel shows that increasing the
interaction parameter $b^2$ drives $v^2_{s}$ toward more negative
values. Next, we turn to the statefinder analysis of the KGDE
model.
\subsection{Statefinder}
A useful tool for distinguishing different cosmological models
from the standard $\Lambda$CDM model is the statefinder
diagnostic. The statefinder parameters were first introduced by
Sahni \cite{Sahni:2002fz} as a pair of dimensionless geometrical
parameters $\{r,s\}$, constructed from the third time derivative
of the scale factor $a(t)$. The standard $\Lambda$CDM model is
represented by the fixed point $\{r,s\}=\{1,0\}$, which provides a
natural reference for comparing various dark energy scenarios. The
statefinder parameters $\{r,s\}$ are explicitly defined as follows
\cite{Sahni:2002fz,Alam:2003sc}
\begin{align}\label{rsdef}
r&=\frac{\dddot{a}}{aH^3}=2\,q^2+q-q^{\prime }, \notag \\
s&=\frac{r-1}{3(q-1/2)}.
\end{align}
From these definitions, it is clear that the statefinder
parameters depend only on the expansion history of the
universe,since they are constructed from the scale factor $a(t)$
and its time derivatives. Using the expression for the
deceleration parameter $q$ and after some straightforward
calculations, the statefinder pair for the KGDE cosmology can be
obtained in the form
\begin{align}
r& =-\frac{1}{\left( 2-\Omega _{D}+\lambda \,\Omega _{D}^{2}/H^{3}\right)
^{3}}\left[ 10\Omega _{D}^{3}\left( 1-3\lambda \,\Omega _{D}/H^{3}\right)
\right.   \notag \\
& -6\Omega _{D}^{2}\left( 4-6b^{2}-5\lambda (1-3b^{2})\,\Omega
_{D}/H^{3}\right) +3\Omega _{D}  \notag \\
& \times \left( 7-24b^{2}+9b^{4}+\lambda (11+18b^{2}-21b^{4})\,\Omega
_{D}/H^{3}\right)   \notag \\
& \left. -4\left( 2-9b^{2}+9b^{4}+9\lambda (1-b^{2})\,\Omega
_{D}/H^{3}\right) \right] ,
\end{align}
and
\begin{align}
s& =-\frac{2\left( 1-b^{2}-\Omega _{D}\right) \left( 2b^{2}+\Omega
_{D}+\lambda (2-2b^{2}-\Omega _{D})\,\Omega _{D}/H^{3}\right) }{\left(
2-\Omega _{D}+\lambda \,\Omega _{D}^{2}/H^{3}\right) ^{2}}  \notag \\
& \times \left[ 4\left( b^{2}-\lambda \,\Omega _{D}/H^{3}\right) +\Omega
_{D}\left( 1-3b^{2}+\lambda (1+7b^{2})\,\,\Omega _{D}/H^{3}\right) \right.
\notag \\
& \left. -\Omega _{D}^{2}\left( 1-3\lambda \,\Omega _{D}/H^{3}\right) \right].
\end{align}
\begin{figure}
\centering
\includegraphics[scale=0.85]{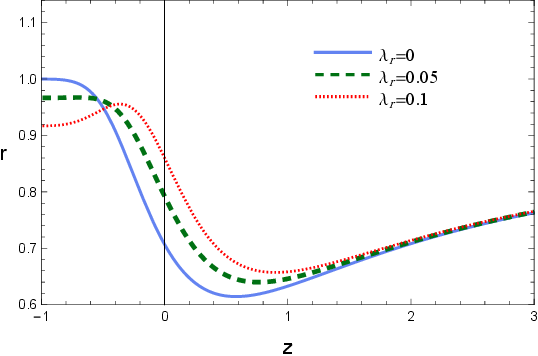}
\includegraphics[scale=0.85]{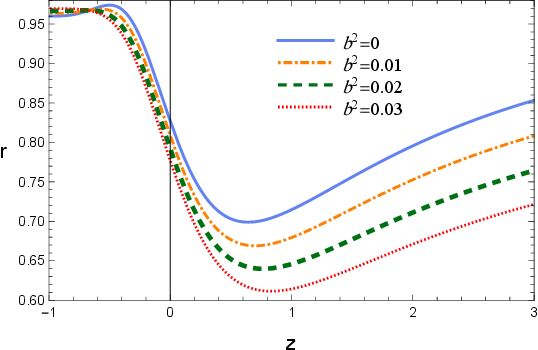}
\vspace{2mm}
\caption{Evolution of $r$ parameter vs $z$ for GDE in Kaniadakis cosmology. In upper
panel we set $b^2 = 0.02$ while in lower panel we take $\lambda_r=0.05$.}\label{fig:r}
\end{figure}
\begin{figure}
\centering
\includegraphics[scale=0.85]{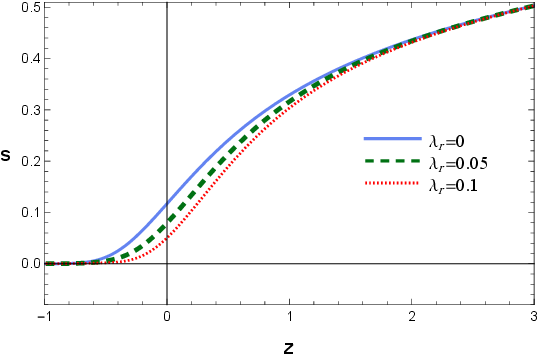}
\includegraphics[scale=0.85]{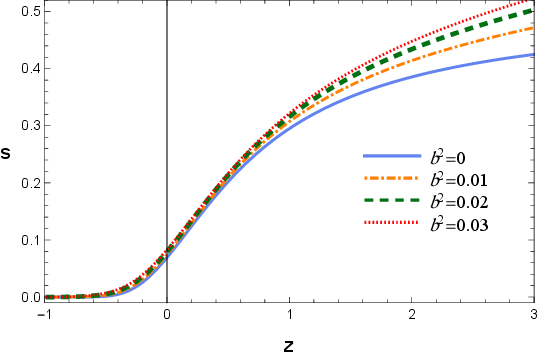}
\vspace{2mm}
\caption{Plot of $S$  vs $z$ for GDE in Kaniadakis cosmology. In upper
panel we set $b^2 = 0.02$ while in lower panel we take $\lambda_r=0.05$.} \label{fig:s}
\end{figure}
In the absence of the Kaniadakis parameter ($\lambda=0$), the
above expressions naturally reduce to the statefinder pair of the
standard GDE model \cite{Malek}. The evolution of the statefinder
parameters $r$ and $s$ for different values of the Kaniadakis
parameter $\lambda_r$ and the interaction parameter $b^2$ is
presented in Figs. \ref{fig:r} and \ref{fig:s}. These figures show
that the KGDE model deviates from the standard $\Lambda$CDM at the
present time. However, for all considered values of $\lambda_r$
and $b^2$, the model approaches the fixed point \{r, s\} = \{1,
0\} in the future. Furthermore, as $\lambda_r$ increases, the
deviation of the model from the standard $\Lambda$CDM scenario
becomes smaller.
\section{Closing remarks}
In this work, we have explored the cosmological implications of
GDE within the framework of Kaniadakis entropy-modified gravity.
Starting from the first law of thermodynamics applied to the
apparent horizon of an FRW universe, we derived modified Friedmann
equations incorporating a Kaniadakis correction term. This
modification naturally gives rise to an effective cosmological
constant, providing a novel link between gravitational
thermodynamics and late-time cosmic acceleration.

We then investigated the dynamics of a spatially flat universe
filled with pressureless matter and interacting GDE in this
modified cosmological setup. By deriving the modified evolution
equation for the GDE density parameter $\Omega_D$, we performed a
detailed numerical analysis to examine the effects of the
Kaniadakis parameter $\lambda$ and the interaction coupling $b^2$
on the cosmological evolution. Our results indicate that while the
Kaniadakis correction has a mild effect on the evolution of
$\Omega_D$, it slightly influences the dark energy EoS parameter,
$w_D$, around the present epoch and shifts the transition redshift
to cosmic acceleration. Notably, the inclusion of the interaction
term allows the GDE model to cross the phantom divide ($w_D = -1$)
at late times.

The stability analysis via the squared sound speed $v_s^2$
revealed that the KGDE model remains generally unstable against
perturbations for the considered parameter ranges. However, we
found that increasing $\lambda$ moderates the instability, while
stronger interactions drive $v_s^2$ toward more negative values.
The statefinder diagnostic $\{r,s\}$ showed that the model
deviates from the standard $\Lambda$CDM paradigm at the present
time but approaches the fixed point $\{1,0\}$ in the future.
Interestingly, larger values of the Kaniadakis parameter lead to
smaller deviations from $\Lambda$CDM, suggesting that entropy
corrections may help reconcile GDE models with observational
constraints.

Future work could extend this analysis by considering more general
interaction forms, studying the perturbation growth and matter
power spectrum, or testing the model against the latest
observational data from cosmic microwave background, baryon
acoustic oscillations, and supernovae surveys. Additionally,
exploring the inflationary implications and the early-universe
dynamics within this entropy-corrected framework would provide
further insights into the viability of Kaniadakis cosmology.
\newpage

\end{document}